\documentclass{article}
\usepackage[utf8]{inputenc}
\usepackage{authblk}
\usepackage{setspace}
\usepackage[margin=1.25in]{geometry}
\usepackage{graphicx}
\graphicspath{ {./figures/} }
\usepackage{subcaption}
\usepackage{amsmath}
\usepackage{lineno}
\usepackage{hyperref}
\usepackage{hyperref}
\usepackage{pgfplots}
\usepackage{tabularx}
\usepackage{booktabs}
\usepackage{verbatim}
\usepackage{listings}
\usepackage{ragged2e}
\usepackage{float}
\usepackage{graphicx}
\usepackage{array}
\newcommand{\keywords}[1]{%
  \begin{center}
    \textbf{Keywords:} #1
  \end{center}
}
\usepackage[style=nejm, 
citestyle=numeric-comp,
sorting=none]{biblatex}
\title{Can Large Language Models Effectively Process and Execute Financial Trading Instructions?}

\author[1,5$\dag$]{Yu Kang}
\author[2,5$\dag$]{Ge Wang}
\author[3,5$\dag$]{Xin Yang}
\author[4,5$\dag$]{Yuda Wang}
\author[5*]{Mingwen Liu}
\affil[1]{Xi'an Jiaotong-Liverpool University}
\affil[2]{The Hong Kong University of Science and Technology}
\affil[3]{Sun Yat-sen University}
\affil[4]{The University of Hong 
Kong}
\affil[5]{Likelihood Lab}
\affil[*]{Address correspondence to: maxwell@xiaochuang.ai}
\affil{Contributing authors:yu.kang23@student.xjtlu.edu.cn; gwangbd@connect.ust.hk; yangx367@mail2.sysu.edu.cn; yuda\_wang@connect.hku.hk}
\affil[$\dag$]{These authors contributed equally to this work.}
 
\date{}

\begin{document}

\maketitle

\begin{abstract}
The development of Large Language Models (LLMs) has created transformative opportunities for the financial industry, especially in the area of financial trading. However, how to integrate LLMs with trading systems has become a challenge. To address this problem, we propose an intelligent trade order recognition pipeline that enables the conversion of trade orders into a standard format in trade execution. The system improves the ability of human traders to interact with trading platforms while addressing the problem of misinformation acquisition in trade execution. In addition, we have created a trade order dataset of 500 pieces of data to simulate real-world trading scenarios. Moreover, we designed several metrics to provide a comprehensive assessment of dataset reliability and the generative power of big models in finance by experimenting with five state-of-the-art LLMs on our dataset. The results indicate that while LLMs demonstrate high generation rates (87.50\% to 98.33\%) and perfect follow-up rates, they face significant challenges in accuracy (5\% to 10\%) and completeness, with high missing rates (14.29\% to 67.29\%). In addition, LLMs tend to over-interrogate, suggesting that large models tend to collect more information, carrying certain challenges for information security.

\end{abstract}

\section{Introduction}\label{introduction}

Artificial intelligence technologies have found increasingly widespread applications in the financial sector recently, demonstrating enormous potential in quantitative trading\cite{WealthGuide2024, RevolutionizingFinanceWithLLMs2024,FinRLFramework2023}. LLMs, as a cutting-edge AI technology, are revolutionizing the way financial analysis and decision-making are conducted. LLMs not only process and generate human-like text but also perform complex analysis\cite{FlamesBenchmark2024}, recognize patterns\cite{theodoridis2006pattern}, and provide support for investment decisions\cite{hackathorn2023large}. In the field of quantitative trading, the application of LLMs has opened up new possibilities, making more sophisticated, accurate, and automated investment strategies a reality\cite{shang_effective_usage_of_llms_2024}. 

Traditional quantitative trading strategies involve the use of computer programs to collect and analyze financial data, select the best-performing investments based on statistical models, and automatically execute trades based on predefined rules\cite{systematic_trading_what_is_it_2024}. While such an approach is effective in enhancing the benefits of efficiency, current quantitative trading systems often encounter difficulties in processing natural language inputs, especially when dealing with complex, ambiguous, or incomplete trading orders, which are prone to a range of problems such as recognition failures and honoring of information. This limitation poses a challenge in bridging the gap between human-generated trading strategies and automated execution systems. In addition, in applications directly connected to trading platforms, the accuracy and speed of trade execution are critical, and thus a balance needs to be struck between improving the accuracy of the model and the efficiency of execution \cite{Fan2024}.

Our research addresses these challenges and makes three significant contributions to the field of LLM-driven trade execution including:

\begin{itemize}
    \item Developd an intelligent trade instruction recognition and execution System that can accurately identify trading requirements, proactively request missing information, and converted natural language instructions into a standardized format for automated execution. 
    
    \item Constructed a novel dataset comprising 500 diverse trading instructions, generated using GPT-4o and verified by human experts. This dataset is enhanced with strategic noise to cover a wide range of real-world scenarios.
    
    \item Designed and implemented a comprehensive set of metrics to evaluate five state-of-the-art LLMs in processing trading instructions, offering an in-depth analysis of their capabilities and limitations in the context of automated trading systems. 
\end{itemize}

The paper is organized as follows: Section \ref{Related Works} reviews related work on the application areas of NLP in finance and automated trading systems. Section \ref{Dataset} describes our dataset creation process and preprocessing methods. Section \ref{Methodology} outlines our experimental approach, including model selection and evaluation metrics. Section \ref{Result} presents our experimental results and analysis. Section \ref{pipeline} shows the complete execution system and the process of user interaction. Section \ref{Conclusion} discusses our findings and their implications. Section \ref{Future Work} discusses the future directions of our research. In addition, all results are validated in our system, which is a simulated financial platform in real time. It can ensure that the experiment results perform effectively and reliably in scenarios that closely mirror real-world applications.


\section{Related Works}\label{Related Works}

Natural Language Processing (NLP) has been widely applied in finance, focusing on key areas such as sentiment analysis \cite{FinancialSentimentAnalysis2019}, market prediction \cite{NewsbasedIntelligentPrediction2023}, and intelligent customer service \cite{NLPTechniquesForAutomating2023}. Early research primarily relied on traditional feature extraction techniques, such as the bag-of-words model and TF-IDF \cite{EffectiveTFIDFModel2023}, combined with machine learning algorithms for tasks like market sentiment analysis and stock prediction \cite{MachineLearningAlgorithms2023}.

With the advent of deep learning, models like Recurrent Neural Networks (RNN)\cite{Sherstinsky2020}and Long Short-Term Memory Networks (LSTM)\cite{Hochreiter1997}have gained popularity in dealing with financial time-series data\cite{Rajpoot2023}which can significantly improve the accuracy of market forecasts\cite{ImprovingStockMarketPrediction2024}. At the same time, such models are also good at capturing long-term dependencies, making them very effective in financial applications where historical data plays a key role.

The introduction of transformer-based LLMs (e.g., BERT\cite{Devlin2018} and GPT\cite{Radford2018}) marked a turning point in the field of financial applications.The core of the application of transformer-based models lies in the self-attention mechanism, which allows them to efficiently capture time-series data in a long-term dependencies. Through its parallel processing capabilities and positional coding, Transformer not only excels at understanding and generating text, but also opens up new opportunities for financial analytics\cite{bianchi2024text} and decision making\cite{StrategicBehaviorOfLLMs2024}. For example, BERT has been applied to analyze financial news and predict stock prices\cite{StockPricePrediction2023}, demonstrating its potential for information extraction and market reaction prediction. Besides, recent research has begun to explore the use of LLMs to generate trading signals and develop automated trading strategies, although this area is still in its infancy\cite{Li2024LargeLanguageModels}.

Automated trading systems have long been a focus of quantitative trading, with most research centered on rule-based strategy generation and high-frequency trading \cite{DeepReinforcementLearningInQuantitativeAlgorithmicTrading2024}. However, these systems typically rely on structured data inputs and have limited capacity to process natural language instructions \cite{DeepLearningForFinancialApplications2020}. Platforms like QuantConnect \cite{quantconnect2024} and Quantopian \cite{quantopian2020} enable traders to implement complex strategies via simplified coding interfaces, yet the requirement for significant programming skills restricts their accessibility for non-technical users. This limitation presents an opportunity for LLMs to bridge the gap, enabling more intuitive interaction with automated trading systems through natural language.

Given the limitations of traditional automated trading systems \cite{Dakalbab2024ArtificialIT}, researchers are now exploring the use of LLMs to extract actionable trading information from unstructured natural language inputs \cite{QuantitativeTrading2023}. This approach holds great promise for bridging the gap between human-generated trading strategies and automated execution systems, thereby making quantitative trading more accessible to non-technical users.

Although significant progress has been made in leveraging LLMs to process financial text, directly converting natural language inputs into executable trading commands remains challenging \cite{Li2021}. Current quantitative trading systems still rely heavily on structured data and lack robust support for natural language inputs \cite{Fan2024}. Additionally, for systems integrated with trading platforms \cite{korczak2024multiagent}, maintaining a balance between execution efficiency and model accuracy is critical, presenting a further challenge that needs to be addressed \cite{PracticalApplicationOfDeepRL2024}.

This study addresses these gaps by investigating how effectively LLMs can recognize and extract trading information from natural language inputs. By converting these inputs into a standardized JSON format and integrating them with trading systems, we facilitate rapid and real-time trade execution. Furthermore, through a comparative analysis of multiple LLM models, we assess their performance in handling various types of trading information, offering both theoretical and practical insights for applying LLMs in automated trading.

\section{Dataset}\label{Dataset}
\subsection{Introduction}
To evaluate the capabilities of various LLMs in identifying and extracting trading information, we constructed a test dataset of 500 items. The dataset was generated using \textbf{GPT-4o} and refined with manual adjustments to simulate a variety of everyday conversations in a trading environment. The idea of using a large model to generate the data was influenced by some recent 
work\cite{wei2023instructiongpt, wang2022self}, where we found that gpt4o understands instructions well and generates compliant data. It includes complete trade execution data, scenarios with missing information, and non-trading related content, addressing the challenge of processing natural language inputs in quantitative trading systems.

Our dataset construction method focuses on enhancing model robustness and generalization by simulating real-world scenarios where trading instructions are often incomplete or noisy. We introduce various types of noise and implement data slicing to ensure diversity, covering a wide range of possible incomplete information scenarios. This approach enables the models to operate effectively across different real-world trading situations.

The table \ref{tab:data-types} below shows the distribution of trade-related and trade-agnostic data in our dataset:

\begin{table}[htbp]
\centering
\caption{The proportion of different data types}
\begin{tabular}{@{}m{5cm}m{3cm}@{}}
\toprule
Type & Percentage \\
\midrule
Trade Instruction & 67.4\% \\
Trade-related Data & 22.4\% \\
Other Data & 10.2\% \\
\bottomrule
\end{tabular}
\label{tab:data-types}
\end{table}

\subsection{Preprocessing}

After constructing data entries with comprehensive trading information, we implemented a preprocessing methodology to enhance the dataset's complexity and representativeness. This involved data augmentation through noise injection and data segmentation via slicing, simulating the diverse variations encountered in real-world applications to improve the dataset's fidelity to actual market conditions.

\textbf{Noise Injection}: We augmented the original data with various linguistic elements, including modal words, ambiguous phrases, diverse punctuation, and code-mixing between Chinese and English. This aimed to make the data less recognizable, challenging language models' ability to extract relevant trading information from noisy inputs.

\textbf{Data Segmentation}: We used a systematic slicing technique to divide the original data into segments of about ten words each, simulating fragmented or incomplete trading information typical in real-time trading environments.

The following example (Table \ref{tab:dataprocess}) illustrates our data processing methodology:

\begin{table}[htbp]
\centering
\caption{Example of Data Preprocessing}
\begin{tabularx}{\textwidth}{@{}lX@{}}
\toprule
Data Type & Example \\
\midrule
Original data & The Skyworth figure in my hand has risen a lot, I decided to take advantage of the good market price, sell all 300 shares in my hand. \\
\addlinespace
Noise data & Oh, Skyworth's stock? It's emmm... risen so much! Selling maybe, uh, all 300 shares, capitalizing on the high price. \\
\addlinespace
Sliced data & The Skyworth figure in my hand has risen a lot. \\
\bottomrule
\end{tabularx}
\label{tab:dataprocess}
\end{table}

As demonstrated in the preceding example, we segment the original comprehensive dataset by omitting certain transaction information elements from the dataset containing complete transaction details, while also incorporating specific modal verbs. Concurrently, we incorporated data from everyday conversations into the dataset to serve as noise. Furthermore, we have developed a JSON format encompassing several key elements, including the trading strategy, ticker symbol, trading type, trading price, and the number of trades. It is important to note that the trading strategy is limited to market orders and limit orders, while the trading type can be classified as either a buy or sell transaction.

\subsection{Data Alignment and Trading Instruction Classification}

In our study, we have undertaken a comprehensive manual alignment of all results, with examples presented in the table below. To facilitate this analysis, we manually generated JSON outputs for 472 out of 500 artificially constructed transaction data points. This process was implemented to enable effective alignment and comparison with responses obtained from various LLMs.

The alignment criteria used in our analysis comprises several essential elements of trading transactions. These include the trading strategy (covering both market and limit orders), the ticker symbol, the type of trade (either buy or sell), the trade volume, and the trade price. It is important to note that when the trading strategy is identified as a market order, no specific price is set; therefore, the trading price is represented as "None" in the output JSON format. Additionally, in cases where user input lacks information relevant to a specific alignment criterion, the corresponding item in the output is marked as "null." Despite these clear definitions, our findings show that LLMs still face challenges in accurately distinguishing between market and limit orders. This ongoing difficulty highlights the complexity of interpreting nuanced financial language, emphasizing the need for further refinement in natural language processing within the financial domain. A more detailed analysis of this issue, including its implications for automated trading systems and potential improvements, will be presented in the evaluation section \ref{Result} of this paper.

In our observations, LLMs frequently struggle to differentiate between limit and market orders. To mitigate this issue, it is critical to provide clear definitions of these order types. \textit{Limit order} is defined as a method of buying or selling stocks at a predetermined price or better, with the caveat that execution is not guaranteed if the stock price fails to reach the specified level. In contrast, \textit{market order} is executed immediately at the best available price, ensuring prompt execution but without guaranteeing a specific price.

This distinction is crucial for the accurate interpretation and processing of trading instructions, particularly in the context of automated systems and algorithmic trading. The ability to correctly identify and handle these different order types has significant implications for trading outcomes and risk management.

\begin{table}[H]
\centering
\caption{Examples of manual alignment results}
\label{tab:dataexample}
\begin{tabular}{m{7cm} m{7cm}} 
\toprule
\textbf{User Input} & \textbf{Json Output} \\ 
\midrule
\centering If Moutai's stock price can fall to 1800, I will take the opportunity to stock up and plan to buy 200 shares of it. & 
\begin{lstlisting}
{
  "strategy": "limit order",
  "symbol": "600519",
  "order_type": "buy",
  "price": 1800.0,
  "quantity": 200
}
\end{lstlisting} \\ 
\midrule
\centering Looking at Vanke stocks to go up, I intend to sell 300 shares first, the bottom position in the hand will not move, and see whether there is room to rise behind & 
\begin{lstlisting}
{
  "strategy": null,
  "symbol": "000002",
  "order_type": "sell",
  "price": null,
  "quantity": 300
}
\end{lstlisting} \\ 
\midrule
\centering I intend to buy 100 shares of Kweichow Moutai while the current stock price is reasonable & 
\begin{lstlisting}
{
  "strategy": "market order",
  "symbol": "600519",
  "order_type": "buy",
  "price": "None",
  "quantity": 100
}
\end{lstlisting} \\ 
\bottomrule
\end{tabular}
\end{table}

\section{Methodology}\label{Methodology}

\subsection{Model}\label{Model}

To conduct our experiment, we selected a range of LLMs that are widely used or high potential. Our selection criteria focused on models with strong instruction-following capabilities, good understanding of Chinese, and accessibility to common researchers. The chosen models include:
\begin{itemize}
    \item \textbf{GPT-4o and GPT-4o-mini \cite{GPT-4omini}}: GPT-4o is renowned for its superior instruction-following and semantic analysis capabilities. We included GPT-4o-mini as a more cost-effective alternative.
    \item \textbf{Qwen-max-0428\cite{Qwen-max-0428}, DeepSeek-v2.5\cite{liu2024deepseek}, and Yi-large\cite{young2024yi}}: These models were selected due to their strong performance with Chinese language content, catering to our target user base of Chinese speakers who may prefer domestic models.
\end{itemize}

This diverse selection allows us to evaluate models with varying strengths in language understanding, instruction following, and domain-specific knowledge, providing a comprehensive assessment of LLM capabilities in processing financial trading instructions for Chinese users.

\subsection{Metric}\label{Metric}
In order to accurately evaluate the performance of different models on our prompts, we will design evaluation metrics from shallow to deep according to the experimental process. At the shallowest level, we only check whether the format and structure of the JSON output are correct (Generation Rate). Further, we will check whether the JSON output is missing information (Missing Rate) and contains erroneous information (Error Rate). Only if the above evaluation models can be executed accurately, it is an accurate JSON result (Accuracy). Accurate JSON results may still be incomplete due to the lack of basic information for quantitative trading execution in user input, hence the model needs to follow up with the user. Therefore, we need to determine whether the model will follow up (Follow-up Rate), and further, we need to determine whether the model asks for too little information (Missed Follow-up Rate) or too much information (Extra Follow-up Rate). 

To test the completion of each step of the model's process, we set the following evaluation metrics in order:

\begin{table}[H]
\centering
\caption{Evaluation Metrics}
\resizebox{\textwidth}{!}{
\begin{tabular}{>{\centering\arraybackslash}m{2.3cm} >{\centering\arraybackslash}m{3cm} >{\raggedright\arraybackslash}m{10.4cm}}
\toprule
\textbf{Metric (\%)} & \textbf{Formula} & \textbf{Definition} \\
\midrule
Generation Rate$^{*}$ & 
\(\frac{\text{JSON Outputs}}{\text{Total Inputs}}\) & 
Evaluate whether the model has generated a JSON in the specified format according to the system prompts for user input.\newline \textbf{- JSON Outputs:} The number of JSON results generated \newline \textbf{- Total Inputs:} The number of user inputs \\
\midrule
Missing Rate\textsuperscript{\#}& 
\(\frac{\text{Missing JSON Outputs}}{\text{JSON Outputs}}\) & 
Evaluate whether the model has omitted input.\newline  \textbf{- Missing JSON Outputs:} The number of JSON results with incomplete input \newline 
\textbf{- JSON Outputs:} The number of JSON results generated \\
\midrule
Error Rate\textsuperscript{\#} & 
\(\frac{\text{Error JSON Outputs}}{\text{JSON Outputs}}\)&
Evaluate whether the model has misunderstood input.\newline  \textbf{- Error JSON Outputs:} The number of JSON with wrong output \newline \textbf{- JSON Outputs:} The number of JSON results generated \\
\midrule
Accuracy$^{*}$ & 
\(\frac{\text{Correct JSON Outputs}}{\text{Total Inputs}}\) &
Evaluate whether the JSON results output by the model are accurate.\newline \textbf{- Correct JSON Outputs:} The number of JSON that correctly understand and complete output of input \newline 
\textbf{- Total Inputs:} The number of user inputs  \\
\midrule
Follow-up Rate$^{*}$ & 
\(\frac{\text{Follow-ups}}{\text{Total Required Follow-ups}}\) &
Evaluate whether the model follows up on incomplete JSON results.\newline\textbf{- Follow-ups:} The number of times that the model makes a follow-up \newline
\textbf{- Total Required Follow-ups:} The number of JSON that lack trade execution information and required follow-up  \\
\midrule
Missed Follow-up Rate\textsuperscript{\#} & 
\(\frac{\text{Missing Follow-ups}}{\text{Total Required Follow-ups}}\) &
Evaluate whether the model has missed follow-up questions. \newline  \textbf{- Missing Follow-ups:} The number of times the model failed to follow up on missing information required for trade execution \newline
\textbf{- Total Required Follow-ups:} The number of JSON that lack trade execution information and required follow-up \\
\midrule
Extra Follow-up Rate\textsuperscript{\#} & 
\(\frac{\text{Extra Follow-ups}}{\text{Total Required Follow-ups}}\) &
Evaluate whether the model has asked wrong follow-up questions. \newline \textbf{- Extra Follow-ups:} The number of unnecessary or incorrect follow-ups taken \newline 
\textbf{- Total Required Follow-ups:} The number of JSON results that lack trade execution information and required follow-up \\
\bottomrule
\end{tabular}
}
\end{table}
\raggedright
\textit{*  lower percentage means better result} \\
\textit{\#  higher percentage means better result} \\
\textbf{Note: All metrics are calculated as percentages (out of 100\%).}

\section{Experiment}\label{Experiment}

\subsection{Experiment Setup}
To evaluate the ability of LLMs to generate executable trading instructions from natural language financial strategies and handle incomplete or ambiguous inputs, we used a setup consisting of 2080ti GPU. The software environment included Python 3.8 with PyTorch 1.9. The Xiaochuang Finance API was used to access real-time and historical financial data, and pre-trained models like GPT-4o, Yi-large, and Qwen-max were accessed. To simulate incomplete strategies, we manually created a dataset of 118 trading strategy descriptions, each with systematically omitted fields. These incomplete strategies are designed to test whether the models can accurately detect the missing elements and generate relevant inquiries to gather the necessary information, ultimately constructing a complete and executable trading strategy.
\subsubsection{Format Generation Evaluation}
Format Generation Evaluation task focuses on testing whether  (LLMs) can transform natural language trading strategies into structured, executable JSON trading instructions that can be directly fed into automated trading systems. This is crucial in financial applications where human-readable strategies must be accurately converted into machine-readable formats to ensure seamless strategy execution.
\subsubsection{Inquiry Capability Evaluation}
Inquiry Capability Evaluation task is designed to test the ability of LLMs to handle incomplete or ambiguous trading strategy descriptions. In real-world trading scenarios, users often provide partial instructions or omit key details, which makes it challenging for automated systems to execute these strategies without clarification. A robust model should be able to detect missing fields in the given strategy and proactively ask follow-up questions to gather the necessary information.
For example, given the initial strategy:
"I want to buy 100 shares of a technology stock."
The strategy lacks critical information such as the specific stock symbol (symbol) and the exact order type (order type). The expected behavior of the model is to identify these missing fields and generate follow-up inquiries such as:
"Which stock are you referring to? Could you specify the stock symbol?"
This ability to engage in interactive questioning not only helps in completing the strategy but also enhances the reliability and accuracy of automated trading systems.

\section{Result}\label{Result}
\subsection{Format Generation Evaluation}

The table compares the performance of five models—Yi-large, Deepseek-v2.5, GPT-4o, GPT-4o-mini, and Qwen-max-0428—in generating structured trading strategies. The key metrics evaluated include Generation Rate, Missing Rate, Correctness, and Accuracy. While GPT-4o achieves the highest Generation Rate at 98.33\%, it also suffers from a high Missing Rate (63.56\%), meaning that it frequently omits critical fields like stock symbols or prices. Yi-large, with a lower Missing Rate of 14.29\%, shows better completeness in generating key components, though its Accuracy remains at 10\%, indicating that manual correction is often required.

\begin{table}[h!]
\centering
\caption{Format Generation Evaluation}
\begin{tabular}{lcccc} 
\toprule
\textbf{Model Name} & \textbf{Generation Rate} & \textbf{Missing Rate} & \textbf{Correctness} & \textbf{Accuracy} \\
\midrule
Yi-large         & 87.50\%  & 14.29\%  & 82.86\%  & 10\% \\
Deepseek-v2.5    & 90.27\%  & 29.86\%  & 77.20\%  & 7.5\% \\
GPT-4o           & 98.33\%  & 63.56\%  & 63.56\%  & 10\% \\
GPT-4o-mini      & 92.50\%  & 56.76\%  & 87.39\%  & 5\% \\
Qwen-max-0428    & 89.17\%  & 67.29\%  & 75.70\%  & 7.5\% \\
\bottomrule
\end{tabular}
\end{table}

\begin{figure}[h!]
    \centering
    \includegraphics[width=0.5\linewidth]{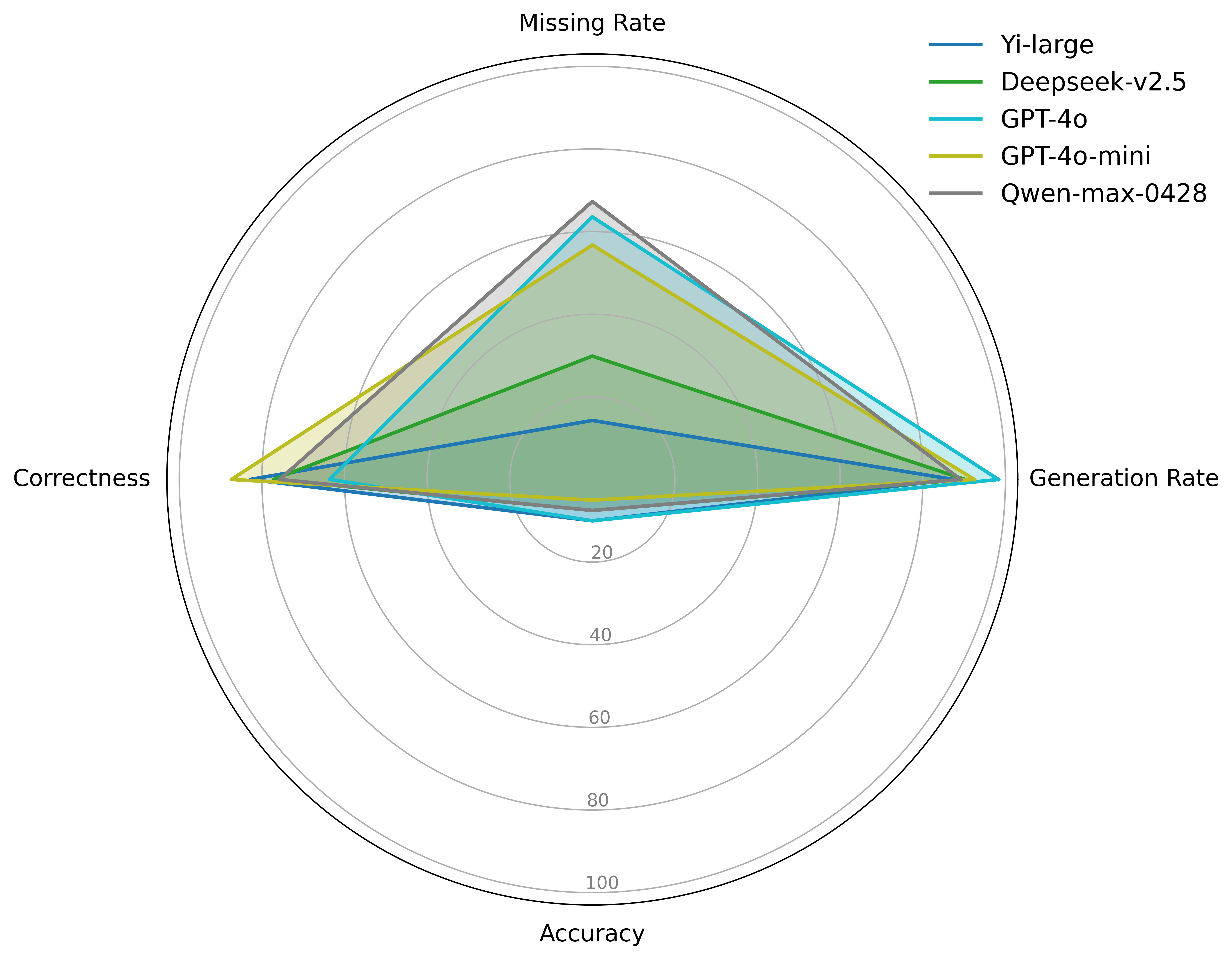}
    \caption{Format Generation Performance Across Models}
    \label{fig:Format Generation}
\end{figure}
The high Missing Rate observed in models like GPT-4o and Qwen-max-0428 (67.29\%) suggests challenges in capturing all essential elements of a trading strategy. These omissions may be due to limitations in domain-specific knowledge or difficulty in handling complex financial terminology. Furthermore, the relatively low Accuracy across all models indicates that while outputs are generated, they often need refinement or adjustment to be executable, particularly when faced with ambiguous or vague input descriptions.

In conclusion, although these models demonstrate potential in generating structured trading strategies, significant improvements are required, especially in reducing the Missing Rate and enhancing Accuracy. To ensure these models are reliable for real-world financial applications, further refinement is necessary to improve their ability to understand and accurately represent complex financial instructions.

\subsection{Inquiry Capability Evaluation}

\begin{table}[h!]
\centering
\caption{Inquiry Capability Evaluation}
\begin{tabular}{lccc} 
\toprule
\textbf{Model Name} & \textbf{Follow-up Rate} & \textbf{Missed Follow-up Rate} & \textbf{Extra Field} \\
\midrule
Yi-large         & 100\%  & 43.22\%  & 76.27\% \\
Deepseek-v2.5    & 100\%  & 29.60\%  & 77.34\% \\
GPT-4o           & 100\%  & 16.95\%  & 91.53\% \\
GPT-4o-mini      & 100\%  & 15.25\%  & 85.59\% \\
Qwen-max-0428    & 100\%  & 27.97\%  & 81.36\% \\
\bottomrule
\end{tabular}
\end{table}

The results indicate that the latest LLM models exhibit a consistent follow-up rate of 100\%, demonstrating their strong capability in generating inquiries when faced with incomplete information. Among the models, \textbf{GPT-4o-mini} and \textbf{GPT-4o} show the lowest missed follow-up rates at 15.25\% and 16.95\%, respectively, suggesting that they are the most reliable at capturing essential information and ensuring the completeness of generated outputs.

\begin{figure}[h!]
    \centering
    \includegraphics[width=0.45\linewidth]{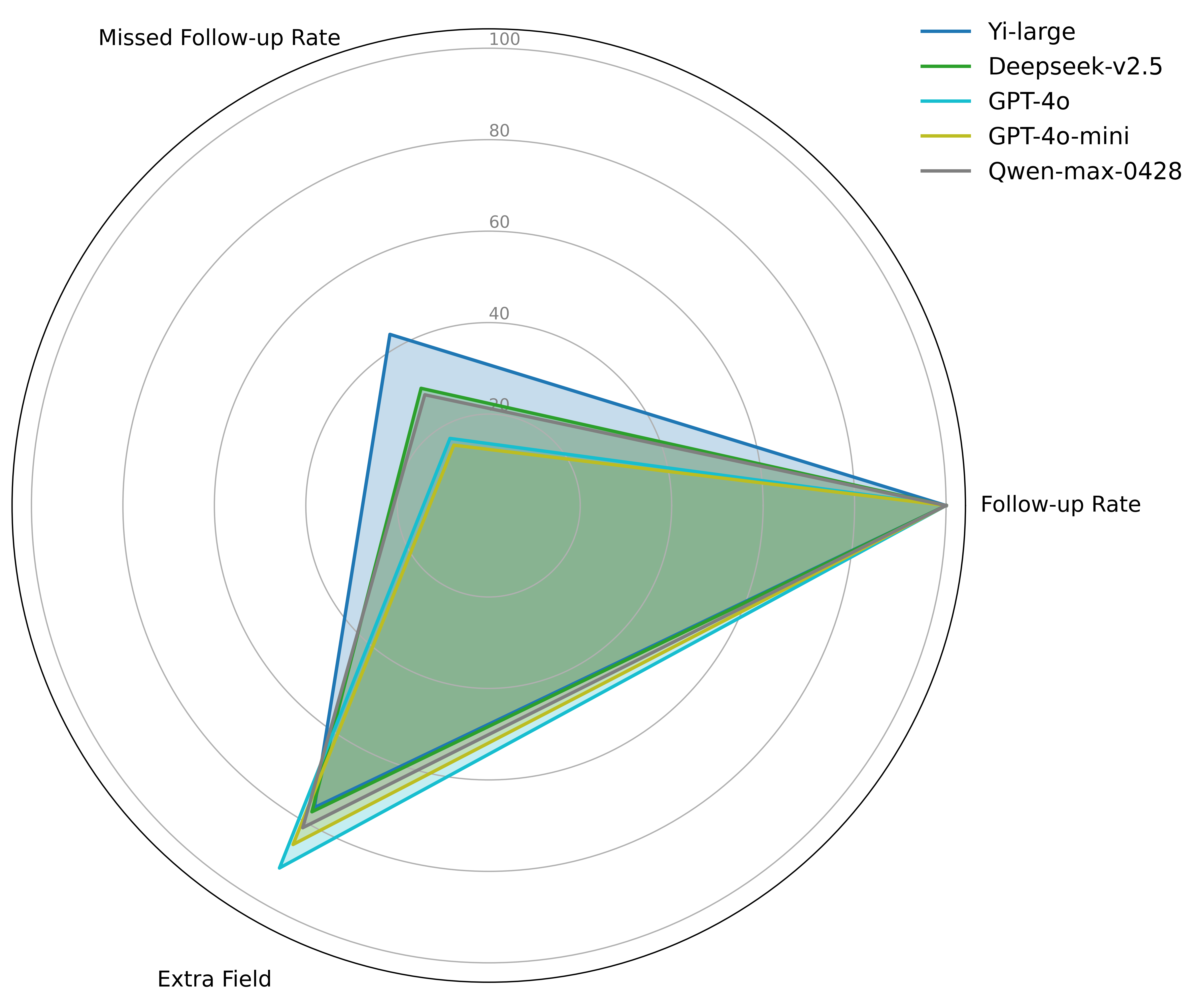}
    \caption{Inquiry Capability Performance Across Models}
    \label{fig:Inquiry Capability}
\end{figure}

Despite the models' success in following up with missing information, all of them still exhibit high \textbf{extra field rates}, ranging from 76.27\% to 91.53\%. This means that the models often ask for or generate information that is not critical to the task at hand. This tendency towards over-specification suggests that, while the models aim to ensure no key details are omitted, they frequently err on the side of caution by including irrelevant or unnecessary elements.

\begin{table}[h!]
\centering
\begin{minipage}{\textwidth}
\caption{Example of Missing and Extra Fields}
\begin{tabular}{p{1.5cm} p{5.5cm} p{5.5cm} p{1.5cm}}
\toprule
\textbf{} & \textbf{Case 1} & \textbf{Case 2} & \textbf{} \\
\midrule
& I want to sell 200 shares of Tencent at \$500 per share. & I intend to buy 100 shares of Kweichow Moutai while the current stock price is reasonable. &  \\
\midrule
\textbf{Cause of Error} & The model correctly identified price and quantity, but added unnecessary follow-up questions. & The price field is vague, leading to incomplete output with missing fields. & \\
\midrule
\textbf{JSON Output} & 
\begin{lstlisting}
{
  "strategy": "market order",
  "symbol": "00700",
  "order_type": "sell",
  "price": 500,
  "quantity": 200
}
\end{lstlisting} & 
\begin{lstlisting}
{
  "strategy": "market order",
  "symbol": "600519",
  "order_type": "buy",
  "price": "None",
  "quantity": 100
}
\end{lstlisting} & \\
\midrule
\textbf{Follow-up Question} & "Would you like to use a market or limit order?" & "What is the price you are willing to pay for Kweichow Moutai?" & \\
\midrule
\textbf{True Question} & No further questions are needed. The price and other details are already provided. & "Please provide the specific price you are willing to pay." & \\
\bottomrule
\end{tabular}
\end{minipage}

\vspace{1em}

\begin{minipage}{\textwidth}
\centering
\includegraphics[width=0.5cm]{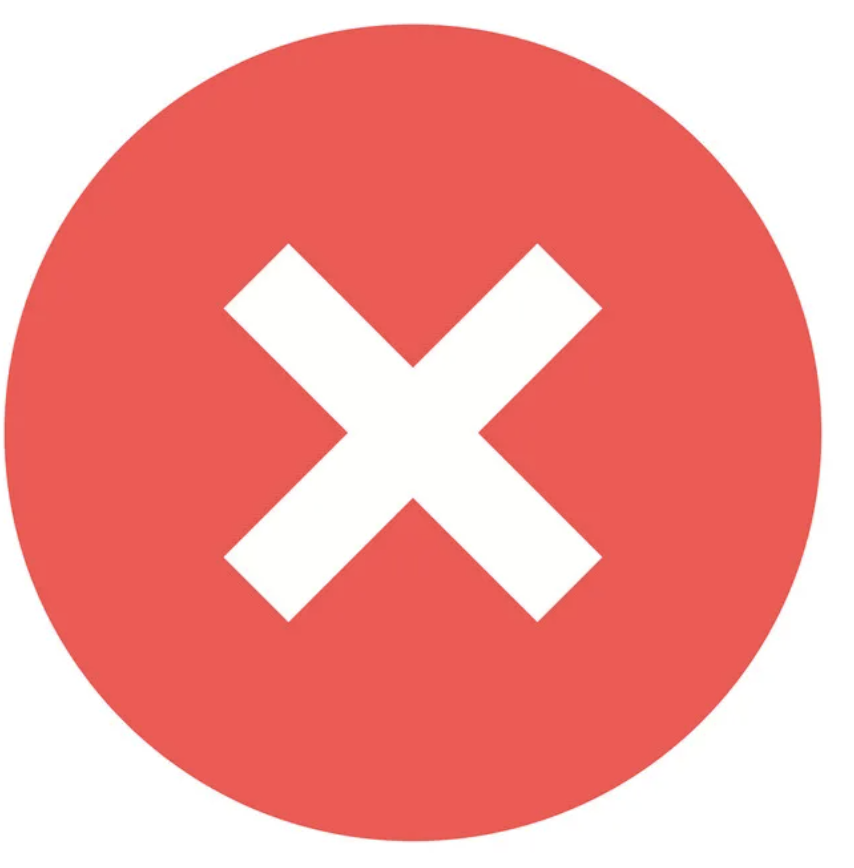} \quad
\textit{Wrong since mis-follow-up} \quad
\includegraphics[width=0.5cm]{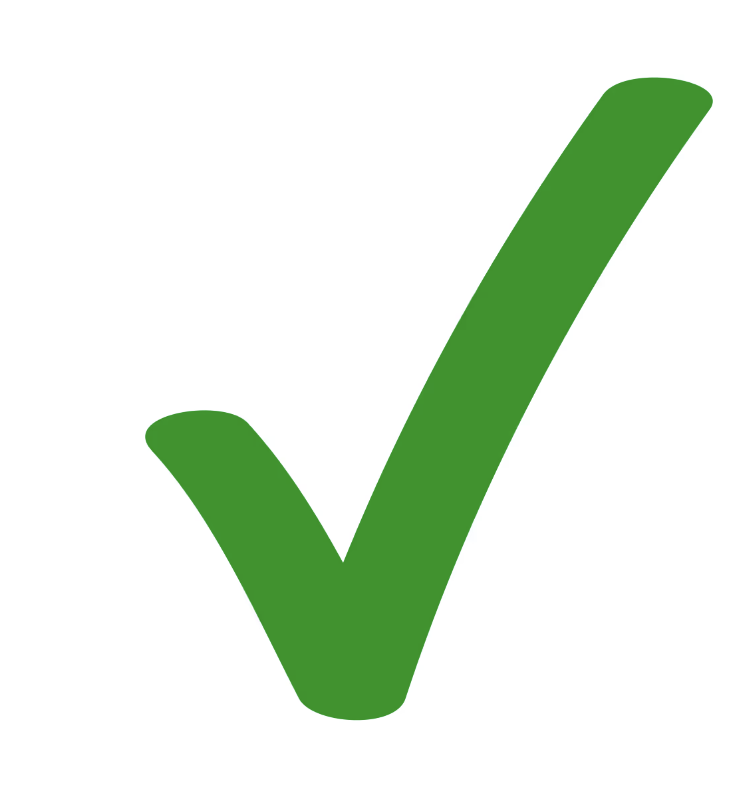} \quad
\textit{Correct follow-up}
\end{minipage}
\end{table}

The cases shown that the user input specifies a straightforward sell order for 200 shares of Tencent at 500 per share. The expected output should be a simple execution of the sell order, with no additional inquiries necessary. However, the model sometimes generates unnecessary follow-up questions, leading to incorrect behavior. For example, GPT-4o might ask:

\begin{quote}
\textit{“Would you like to use a market or limit order?”}
\end{quote}

This follow-up question is irrelevant since the user has already provided sufficient details, including the order type (sell) and price. Therefore, no further clarification should be required.

On the other hand, in the second example where the user plans to buy 100 shares of Kweichow Moutai, the price is not specified, leading to a missing field in the generated output. In this case, the model appropriately asks:

\begin{quote}
\textit{“What is the price you are willing to pay for Kweichow Moutai?”}
\end{quote}

This is the correct behavior, as the input was incomplete and the price field was missing. The follow-up question accurately seeks to gather the necessary information to execute the order properly.

\textbf{Yi-large}, despite having the highest missed follow-up rate of 43.22\%, compensates by asking fewer unnecessary questions, reflected in its lower extra field rate of 76.27\%. This more balanced approach makes it more reliable in scenarios where over-inquiry could lead to confusion or inefficiency. While Yi-large occasionally misses important details, its more conservative approach to generating extra fields makes it more efficient in certain contexts, where brevity and clarity are essential.

\textbf{Qwen-max-0428}, with a missed follow-up rate of 27.97\% and an extra field rate of 81.36\%, demonstrates balanced performance but could still benefit from refinement. Notably, \textbf{GPT-4o} has shown high scores in recent financial strategy generation leaderboards, such as the \textit{Hugging Face Financial Leaderboard} and \textit{Financial NLP Benchmark}, reflecting its overall robustness in handling complex, domain-specific queries. Despite this, its relatively high missed follow-up rate (16.95\%) and extra field rate (91.53\%) suggest that it could still benefit from improvements in better distinguishing essential from non-essential elements, ensuring higher precision in generating complete, yet concise, strategies.This analysis reflects the trade-offs between completeness and over-inquiry, with models becoming more reliable in retaining key elements, but at the cost of generating extra, sometimes unnecessary, information.

Besides, in our experiments, we observed three significant challenges: The LLMs demonstrated a limited understanding of trading knowledgeand had difficultly in properly matching specialized financial terms. This limitation might be attributed to the non-specialized general domain corpus used to train the models. The specific meanings of trading terms, often distinct from their everyday usage, presented a significant challenge for the models. Secondly, while the models exhibited proficiency in generating JSON structures, they consistently failed to distinguish between string representations (e.g., "NULL") and actual null values, despite our explicit instructions and emphasis on this distinction. This issue suggests that large models have significant problems understanding JSON syntax and semantics, especially in terms of data representation. Thirdly, in our JSON follow-up experiments, the model lacks the ability to identify missing data items without explicitly reflecting the steps.

\section{Execution Pipeline}\label{pipeline}

\begin{figure}[H]
    \centering
    \includegraphics[width=1.0\linewidth]{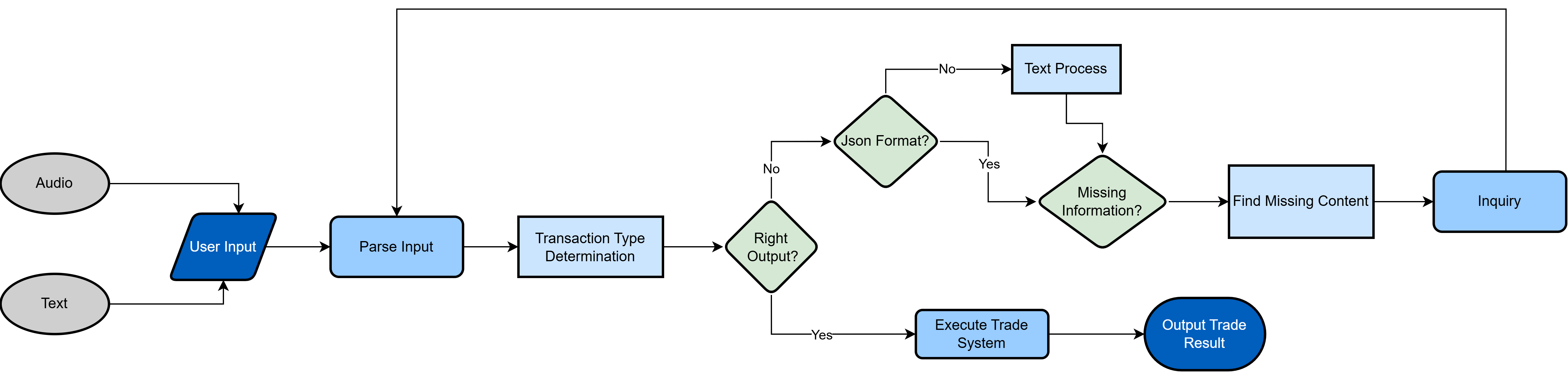}
    \caption{Execution Pipeline}
    \label{fig:pipleine}
\end{figure}

In order to address some of the mentioned issues above (e.g., incorrect JSON-generated content), we developed a comprehensive transaction execution process. This ensures that our transactions are fully executed, comprising everything from user input and parsing to the execution during the pipeline. Figure \ref{fig:pipleine} illustrates how we accomplish this. Initially, the system receives input from the user, which may be in either voice or text format. Next, the system must parse this input to effectively process and comprehend the user's request. Following the parsing phase, the system proceeds to the transaction type determination stage, which primarily involves identifying whether the transaction is a \textbf{Market Order} or \textbf{Limit Order}. In the case of a market order, the price is set to default as empty. This step is critical, as prior evaluations have indicated that a failure to recognize the type of transaction significantly diminishes the accuracy of the final result.

The subsequent step involves the system's assessment of its capability to accurately produce the transaction results. This represents a critical decision-making juncture within the process. Should the system successfully generate the transaction output, it will proceed directly to the execution phase of the transaction. During this execution phase, the system conducts the actual operations within the trading framework following the user's request, culminating in the completion of the transaction and the display of results. If the system is unable  to generate a valid output, it will transition to a specialized handling the step. This phase is entirely governed by the overarching model, beginning with the system's verification of whether the data output adheres to the JSON format. If the output data is indeed in JSON format, the system will engage in an internal processing workflow to further parse and provide feedback on the data. In instances where the data does not conform to the JSON format, the system will initiate a text processing workflow for additional data handling. Once the data format as JSON has established, we can determine if any loss occurs when processing. If there is no loss of data, the system will continue to the subsequent operations successfully; however, if it identifies any missing content, it will prompt the user until all necessary information is provided. This iterative process ensures the completeness and accuracy of the transaction data before final execution. The design of this pipeline addresses the challenges identified in our experiments, particularly in recognizing financial terminology, distinguishing between JSON string representations and special values, and detecting missing data. By incorporating these improvements, the system aims to significantly enhance the efficiency and reliability of automated trading instruction processing, paving the way for more intelligent financial systems in the future.We provide an real execution example shown below:

\begin{figure}[H]
    \centering
    \includegraphics[width=0.8\linewidth]{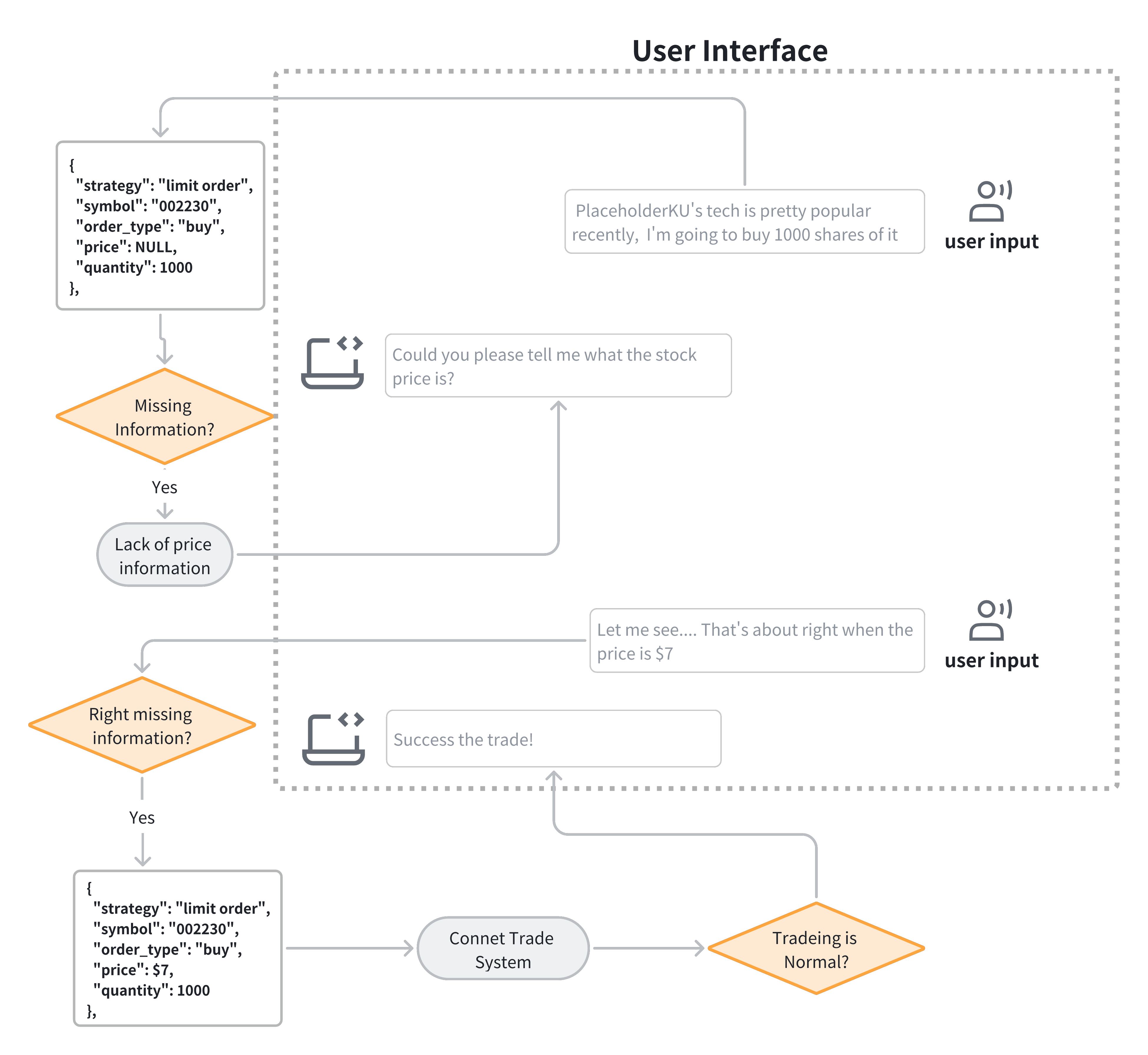}
    \caption{User Interaction}
    \label{fig:interaction}
\end{figure}

The system(Fig \ref{fig:interaction}) is mainly a intelligent trading instruction processing so the user interaction figure denotes that it shows working of UI which processes such instructions to be processed directly due for only UX purpose. This is a complex step process, starting from the user input and ending with executing a trade. The user starts with an input, like this: "KU's tech is going to be really popular, I'll buy 1000 shares of it." The system finally processes this input and gives as a part of JSON structure which consists important trading parameters such strategy, symbol, order\_type, price,quantity etc. An important property illustrated in this figure is asking for absent information from the system. As an example when the system identifies that a limit order is without price information, it asks this way:, "Could you please tell me what the stock price is?" This interaction ensures that all necessary details will be collected before entering trade system. The figure also shows how the system handles the user's response, updating the JSON output with the provided information (in this case, setting the price to \$7). Therefore, once all required information is gathered and verified, the system proceeds to execute the trade, confirming the success with a message to the user.

\section{Conclusion}\label{Conclusion}
In this study, we investigate the capabilities of LLMs in processing and executing financial trading orders. We developed a comprehensive evaluation framework and conducted experiments on five state-of-the-art LLMs. The results indicate that LLMs excel in generating structured outputs, with generation rates ranging from 87.50\% to 98.33\%. However, they still encounter significant challenges regarding accuracy and completeness, with missing rates between 14.29\% and 67.29\%, and accuracies ranging from a mere 5\% to 10\%. Our proposed novel pipeline integrates LLMs with additional processing steps, demonstrating the potential to enhance the performance of LLMs in financial applications. Nevertheless, the experiments also highlight the limitations of LLMs in recognizing discourse-level incoherence, inference knowledge, generalization ability, and robustness. 

\section{Future Work}\label{Future Work}

Our results provide a preliminary demonstration of the advantages and disadvantages associated with LLMs in the field of financial transactions. In the future, we intend to make improvements in the following areas:

1. Enhancing the Dataset: Our current dataset, although informative, is limited in both quantity and quality. In future work, we plan to expand the data set to more than 3,000 entries, significantly increasing its size and representativeness. We will focus on enriching the diversity of trading scenarios, and providing a more comprehensive coverage of financial instruments and trading strategies.

2. Pipeline Optimization: We aim to improve the proposed pipeline to increase its accuracy and efficiency within a real-time trading environment. A key improvement will involve the integration of real-time market data sources. Furthermore, we plan to implement a risk assessment and early warning system that will evaluate the potential risks associated with trade order before execution. This functionality will help minimize potential losses and provide users with valuable, real-time insights into their trading decisions.

3. Exploring the Boundaries of LLMs in Finance: We intend to investigate the performance of these models in more complex financial tasks, such as portfolio optimization, risk management, and market sentiment analysis.

\section{Acknowledge}

This research was supported by Likelihood Lab\footnote{http://maxlikelihood.cn/} and Xiaochuang Intelligence\footnote{https://xiaochuang.ai/}. All APIs used in the study were generously provided by Xiaochuang Intelligence. We express our sincere gratitude for their invaluable support.

\printbibliography

\end{document}